\documentclass[aps,pre,reprint,superscriptaddress,showkeys,showpacs]{revtex4-2}

\usepackage[utf8]{inputenc}
\usepackage[T1]{fontenc}
\usepackage[english]{babel}
\usepackage{stmaryrd} 

\usepackage{amsmath, amssymb, amsthm, amsfonts}
\usepackage{mathrsfs}        
\usepackage{physics}         
\usepackage{graphicx}        
\usepackage{ifpdf}
\ifpdf
 \DeclareGraphicsExtensions{.pdf,.eps}
\else
 \DeclareGraphicsExtensions{.eps}
\fi
\usepackage{xcolor}          
\usepackage{caption}
\usepackage{subcaption}      
\usepackage{ragged2e}
\usepackage{array,booktabs}  
\DeclareCaptionJustification{fulljust}{\justifying}
\usepackage{geometry}
\usepackage{setspace}        
\usepackage{csquotes}        
\usepackage{enumitem}
\usepackage{multido} 
\ifpdf
 \usepackage[breaklinks=true]{hyperref}
\else
 \providecommand{\texorpdfstring}[2]{#1}
\fi

\makeatletter
\renewcommand{\@pacs@name}{2020 Mathematics Subject Classification: }
\makeatother
\newcommand{\subjclass}[1]{\pacs{#1}}
\newcommand{\articlerunningtitle}{Stress overshoots and stick-slip in mean-field depinning}

\theoremstyle{remark}

\newcommand{\barv}{\bar{v}}

\begin{document}
\title[\articlerunningtitle]{Velocity weakening and stick-slip in mean-field depinning with stress overshoots}

\author{Adrien Rezzouk} 
\affiliation{LPTMS, Université Paris Saclay, CNRS, 91405 Orsay, France}
\author{Alberto Rosso}
\affiliation{LPTMS, Université Paris Saclay, CNRS, 91405 Orsay, France}

\begin{abstract}
{We study a fully connected depinning model with transient
stress overshoots of strength $M$, driven through a spring of stiffness
$k_0$.  Combining
exact stationary relations, linear stability analysis, direct simulations,
and population dynamics, we connect the force--velocity characteristic to
stick-slip motion and quasistatic avalanche statistics. For $M>1$, two distinct
stiffness scales organize the response.  At
$k_0=k_{\rm w}\equiv M-1$, the stationary
characteristic first develops a velocity-weakening branch, whereas
$k_0=k_{\rm q}\equiv(M-1)^2/(4M)$ marks the onset of finite-amplitude stick-slip in
the quasistatic limit.  Their separation produces an intermediate regime in
which the quasistatic force converges to a stationary value, whereas
finite-velocity driving produces stick-slip below a critical velocity.
Quasistatic avalanches in this regime undergo transient
amplification and display an excess of large, asymmetric events---
``ghost kings''---that remain finite. By contrast, at and below $k_{\rm q}$ the same
amplification produces system-spanning ``genuine kings'' and a finite
quasistatic force cycle.  These results show how elastic loading dynamically selects among the
branches of a reentrant force--velocity relation and clarify the connection
between velocity weakening, stick-slip, and runaway avalanches.}
\end{abstract}
\maketitle
\markboth{A. Rezzouk and A. Rosso}{\articlerunningtitle}

\begingroup
Driven elastic systems in random media display a depinning transition between
a pinned phase and a steadily moving phase.  In the overdamped limit this
transition is continuous: the mean velocity vanishes at a critical force,
while quasistatic loading produces scale-free avalanches whose cutoff diverges
on approaching depinning
\cite{NarayanFisher1992,NattermannStepanowTangLeschhorn1992,Fisher1998,Wiese2022}.
This framework describes intermittent motion in magnetic domain walls
\cite{GrassiKoltonEtAl2018}, contact lines
\cite{MoulinetRossoKrauthRolley2004}, crack fronts
\cite{MaloySantucciSchmittbuhlToussaint2006,BonamySantucciPonson2008}, and
fault models
\cite{FisherDahmenRamanathanBenZion1997,PetrilloLippielloLandesRosso2020}.
The overdamped description, however, neglects dynamical mechanisms such as
inertia, delayed stress transfer, and internal relaxation.  Although
microscopically distinct, these mechanisms can transiently sustain the motion
of regions that have already started to slip.

Fisher et al. introduced a fault model in which motion transiently enhances
the stress transferred to neighboring regions
\cite{FisherDahmenRamanathanBenZion1997}. Schwarz and Fisher subsequently
analyzed the corresponding mean-field depinning model
\cite{SchwarzFisher2001}. They showed that transient stress overshoots
can turn a continuous depinning transition into a discontinuous one,
producing a reentrant moving branch, force hysteresis, and coexistence between
critical avalanches and characteristic large events.  In finite
dimensions the associated hysteresis can become pseudohysteretic and vanish
asymptotically even though broad crossover regimes remain
\cite{SchwarzFisher2003}.  Related departures from overdamped depinning arise
from viscoelasticity and plastic relaxation
\cite{MarchettiMiddletonPrellberg2000,MarchettiMiddletonSaundersSchwarz2003,
Marchetti2006}.

Dynamical effects beyond the overdamped limit have also been studied in
spatial models. Inertia and weak damping can strongly modify avalanche
statistics in amorphous solids
\cite{SalernoRobbins2013,KarimiFerreroBarrat2017}. Slow internal relaxation
generates velocity weakening and clustered events in spring--block systems
\cite{JaglaKolton2010}, as well as quasiperiodic avalanche cycles in crystal
plasticity~\cite{PapanikolaouEtAl2012}. Viscoelastic depinning likewise
produces nonstationary avalanche sequences and stress oscillations
\cite{JaglaLandesRosso2014}.

A central issue is how these dynamical instabilities are selected by the
driving protocol. When the system is loaded through a spring, rather than at
fixed force, the force itself becomes a dynamical variable: it increases
between events and decreases during slip. The spring stiffness can therefore
determine whether an instability develops into steady motion, stick-slip, or
an arrested avalanche. More recently, a study of a finite-dimensional model
beyond the overdamped limit has distinguished genuine system-spanning avalanches from
\emph{ghost kings}: supercritical events whose characteristic size remains
finite in the thermodynamic limit. In short-range systems, these events
propagate ballistically~\cite{PetrilloJaglaLippielloRosso2026}.
Related arrest-versus-runaway questions arise in amorphous solids, where the
nucleation of a shear band can produce discontinuous failure
\cite{OzawaBerthierBiroliRossoTarjus2018,PopovicDeGeusWyart2018}, and at
frictional interfaces, where collective asperity detachments can nucleate
macroscopic slip
\cite{deGeusPopovicJiRossoWyart2019,deGeusWyart2022}.

Here we revisit the fully connected stress-overshoot model of Schwarz and
Fisher, in which $M$ controls the strength of the transient stress overshoot.
We drive the system through a spring of stiffness $k_0$. For $M>1$, as $k_0$ is
decreased, velocity weakening first appears at
$k_0=k_{\rm w}\equiv M-1$, whereas quasistatic stick-slip sets in only at the
lower threshold $k_0=k_{\rm q}\equiv(M-1)^2/(4M)$.  Between these two
thresholds, oscillations exist at finite driving velocity but disappear as
$V\to0$. Quasistatic loading nevertheless produces large avalanches with an
asymmetric shape and a transiently amplified activity; these events are
arrested before becoming system spanning, and we identify them as
\emph{ghost kings}. At and below $k_{\rm q}$, the amplification instead produces
system-spanning avalanches---\emph{genuine kings}. We establish this picture
using exact stationary relations, stability analysis, finite-$N$ simulations,
and population dynamics.

The paper is organized as follows. Section~I introduces the model, and
Sec.~II recalls the force-controlled solution. Section~III presents the
elastic-driving results, from the stationary characteristic to finite-velocity
dynamics and quasistatic avalanches. Section~IV discusses the resulting
three-regime picture, while numerical details are collected in the Appendices.
\endgroup

\section{Model}
We represent the interface by $N$ blocks with scalar positions
$h_1,\ldots,h_N$. In the overdamped limit, the instantaneous velocity of block
$i$ is the sum of the elastic, pinning, and driving forces,
\begin{equation}
v_i = F_i^{\mathrm{elast}} + F_i^{\mathrm{dis}} + F_i^{\mathrm{drive}} .
\end{equation}
We consider a fully connected elastic interaction,
\begin{equation}
F_i^{\mathrm{elast}} = h_{\mathrm{CM}} - h_i,
\end{equation}
where
\begin{equation}
h_{\mathrm{CM}} = \frac{1}{N} \sum_i h_i
\end{equation}
is the center-of-mass position of the interface.

\begin{figure}[t]
	\centering
	\includegraphics{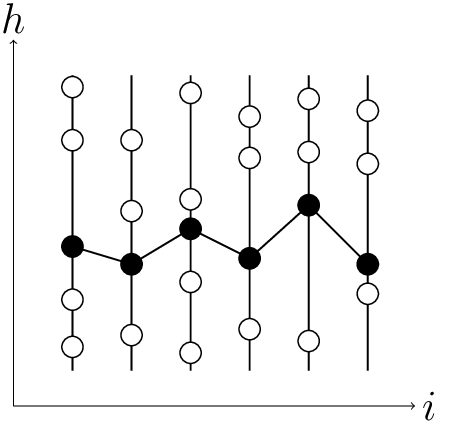}
     \caption{Schematic illustration of the narrow-wells approximation. Open
     circles represent the sticky patches; solid black circles mark the pinned
     block positions.}
    \label{fig:wells}
\end{figure}

Quenched disorder is described within the narrow-wells approximation. For
each block, the disordered landscape is reduced to an independent random
sequence of narrow sticky patches along the $h$ axis, represented by the open
circles in Fig.~\ref{fig:wells}. While a block is trapped at one of these
patches, the pinning force $F_i^{\mathrm{dis}}$ acts as a static-friction
force: it balances the other forces and keeps the block immobile as long as
the local force remains below the yield threshold $f_Y$. When this threshold
is reached, the block becomes unstable and jumps forward to the next sticky
patch. We set $f_Y=1$, thereby fixing the force scale. The distance $\Delta$
between consecutive patches is random; throughout this work, we use the
exponential distribution $g(\Delta)=e^{-\Delta}$, whose unit mean fixes the
length scale. The sequences of sticky patches are statistically independent
for different blocks and homogeneous along the $h$ direction. Consequently,
disorder-averaged quantities are invariant under a uniform translation of
the interface, a property that will be used below when deriving the
stationary force--velocity relation.

We consider two driving protocols. Under force control, each block is
subjected to the same constant external force,
\begin{equation}
F_i^{\mathrm{drive}} = F,
\end{equation}
as in Ref.~\cite{SchwarzFisher2001}. Under elastic driving, each block is
coupled through a spring of stiffness $k_0$ to a common driving position $w$,
\begin{equation}
F_i^{\mathrm{drive}} = k_0 (w - h_i).
\end{equation}
The force averaged over the interface is therefore
\begin{equation}
F_{\mathrm{av}}(w) = \frac{1}{N}\sum_iF_i^{\mathrm{drive}}
= k_0 (w - h_{\mathrm{CM}}).
\end{equation}
We denote by $V=\dot w$ the imposed driving velocity and by
$\bar v=\dot h_{\mathrm{CM}}$ the mean velocity of the interface. These two
velocities are generally distinct during a transient or an avalanche. They
coincide, $\bar v=V$, only in a stationary moving state. In the quasistatic
protocol, $w$ is held fixed during each avalanche and is increased again only
after the entire system has repinned.

We first formulate the microscopic dynamics under force control. The state of
block $i$ is characterized by its \emph{distance to instability},
\begin{equation}
x_i = f_Y - \big(h_{\mathrm{CM}} - h_i\big) - F.
\end{equation}
The block is stable for $x_i>0$ and becomes unstable for $x_i\leq0$.
Consider a forward jump of block $i$ by a distance $\Delta$, while all other
blocks remain pinned. Denoting by a prime the configuration immediately after
this jump, we have
\begin{align}
h_i' &= h_i + \Delta, \\
h_{\mathrm{CM}}' &= h_{\mathrm{CM}} + \frac{\Delta}{N}.
\end{align}
The corresponding distances to instability are
\begin{align}
x_i' &= x_i + \Delta \left(1 - \frac{1}{N}\right), \\
x_j' &= x_j - \frac{\Delta}{N}, \qquad j\neq i.
\end{align}
The jump increases the distance to instability of block $i$, while bringing every other block
closer to instability by $\Delta/N$. Some of these blocks may consequently
become unstable and trigger further jumps, producing an avalanche.

The Schwarz--Fisher model adds a transient elastic overshoot of strength $M$.
Time is now discretized into parallel update steps. The distance to
instability of block $i$ at time $t$ is
\begin{equation}
    x_i(t) = f_Y-[h_{\mathrm{CM}}(t)-h_i(t)]-F-M\barv(t),
    \label{eq:xiM}
\end{equation}
where
\begin{equation}
    \barv(t) = h_{\mathrm{CM}}(t) - h_{\mathrm{CM}}(t-1)
\end{equation}
is the mean displacement of the interface during the preceding update. At
step $t$, all blocks with $x_i(t)\leq0$ jump simultaneously. Denoting their
forward displacements by
\begin{equation}
    \Delta_i(t)=h_i(t+1)-h_i(t),
\end{equation}
with $\Delta_i(t)=0$ for the stable blocks, the mean displacement generated
by this update is
\begin{equation}
    \barv(t+1)=\frac{1}{N}\sum_i\Delta_i(t).
\end{equation}
The distances to instability then evolve according to
\begin{equation}
    x_i(t+1)=x_i(t)+\Delta_i(t)
    +M\barv(t)-(1+M)\barv(t+1).
\end{equation}
The term proportional to $M$ is transient: the activity generated during one
update increases the force acting on all blocks during the next update, but
this additional force is removed one time step later. It can therefore
destabilize further blocks and amplify the propagation of an avalanche
without modifying the static force balance of a pinned configuration.

\section{Constant force and force--velocity characteristics}
We first recover the force--velocity characteristic obtained by Schwarz and
Fisher under force-controlled driving~\cite{SchwarzFisher2001}. We work in
the thermodynamic limit, $N\to\infty$, and impose a constant force $F$.

Throughout the paper, an overbar denotes an average over the quenched
disorder. In the thermodynamic limit, this coincides with the average over
blocks at a fixed macroscopic state. For jump lengths drawn from
$g(\Delta)$ and observables of the distance to instability, respectively,
\begin{align*}
 \overline{\Delta^n}&=\int_0^\infty\Delta^n g(\Delta)\,\mathrm d\Delta,
 \qquad n=1,2,\\
 \overline{B(x)}&=\int_{-\infty}^{\infty}B(x)\rho(x)\,\mathrm dx,
\end{align*}
where $\rho(x)$ is the normalized population density of distances to
instability. A subscript specifies conditioning: the disorder average of
$x$ over stable blocks is
\[
 \overline{x}_{\,x>0}
 \equiv\frac{\overline{x\Theta(x)}}{\mathbb P(x>0)}
 =\frac{\int_0^\infty x\rho(x)\,\mathrm dx}
 {\int_0^\infty\rho(x)\,\mathrm dx}.
\]
Here $\Theta(x)$ is the Heaviside step function: $\Theta(x)=1$ for $x>0$
and $\Theta(x)=0$ for $x<0$. Its value at the origin does not affect the
integrals for the continuous densities considered here.
\begin{figure}[t]
\centering
\includegraphics{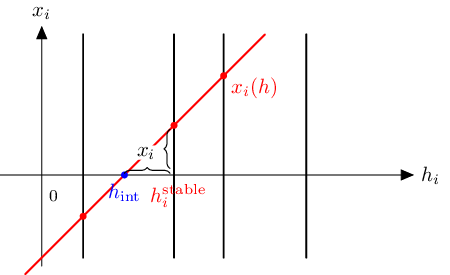}
\caption{Vertical lines mark the sticky-patch positions. The intersection
point $h_{\mathrm{int}}$ separates the unstable region ($x_i<0$) from the
stable region ($x_i>0$). The position $h_i^{\mathrm{stable}}$ is the first
sticky patch in the stable region, and
$x_i=h_i^{\mathrm{stable}}-h_{\mathrm{int}}$.}
\label{fig:xi_pieges}
\end{figure}

For a stationary state with mean velocity $\bar v$, including a pinned state
with $\bar v=0$, Eq.~\eqref{eq:xiM} gives the distance to instability of a
block occupying a patch at position $h$ as
\begin{equation}
 x(h)=h-h_{\mathrm{int}},
 \qquad
 h_{\mathrm{int}}=h_{\mathrm{CM}}+F+M\bar v-f_Y,
 \label{eq:hint}
\end{equation}
where $h_{\mathrm{int}}$ is the position at which
$x(h_{\mathrm{int}})=0$. Sticky patches with $h>h_{\mathrm{int}}$ are stable,
whereas those with $h\leq h_{\mathrm{int}}$ are unstable. We first consider
a pinned reference population in which each block occupies the first sticky
patch to the right of $h_{\mathrm{int}}$, denoted by
$h_i^{\mathrm{stable}}$. Its distance to instability is
\begin{equation}
 x_i=h_i^{\mathrm{stable}}-h_{\mathrm{int}}>0.
\end{equation}

Because the sequences of sticky patches are statistically homogeneous along
$h$, the position of $h_{\mathrm{int}}$ relative to a given sequence is
uniformly distributed. It falls within an interval of length $\Delta$ with
the length-biased probability density
\begin{equation}
 \frac{\Delta g(\Delta)}{\overline{\Delta}}.
\end{equation}
Conditioned on an interval of length $\Delta$, $h_{\mathrm{int}}$ is
uniformly located inside it. Its mean distance from the sticky patch at the
right end is therefore $\Delta/2$. Averaging over the interval lengths gives
the mean distance to instability of a pinned block,
\begin{equation}
 \overline{x}_{\,x>0}
 =\int_0^\infty
 \frac{\Delta g(\Delta)}{\overline{\Delta}}\frac{\Delta}{2}\,\mathrm d\Delta
 =\frac{\overline{\Delta^2}}{2\overline{\Delta}}.
 \label{eq:mean-pinned-distance}
\end{equation}

\paragraph{Derivation of $F_{c0}$.}
As a first application of the previous result, we derive $F_{c0}$, defined as
the minimum force above which no pinned stationary state exists for $M=0$.
At $F=F_{c0}$, all blocks are pinned at their first stable sticky patch
$h_i^{\mathrm{stable}}$. Consequently,
\begin{equation}
 h_{\mathrm{CM}}=\overline{h_i^{\mathrm{stable}}}
 =h_{\mathrm{int}}+\overline{x}_{\,x>0}.
\end{equation}
Since $\bar v=0$, Eq.~\eqref{eq:hint} gives
$h_{\mathrm{int}}=h_{\mathrm{CM}}+F_{c0}-f_Y$. Combining these relations
with Eq.~\eqref{eq:mean-pinned-distance} yields
\begin{equation}
 F_{c0}=f_Y-\frac{\overline{\Delta^2}}{2\overline{\Delta}}.
 \label{eq:Fc0}
\end{equation}

\paragraph{Derivation of the force--velocity equation.}
In a stationary moving state, each block is shifted by $-\bar v$ during one
update, while an unstable block additionally draws a forward jump $\Delta$
from $g(\Delta)$. The stationary density therefore obeys
\begin{align}
 \rho(x)={}&\rho(x+\bar v)\Theta(x+\bar v) \nonumber\\
 &+\int_0^\infty g(\Delta)\rho(x+\bar v-\Delta)
 \Theta(-x-\bar v+\Delta)\,\mathrm d\Delta.
 \label{eq:stationary-master}
\end{align}
The transient overshoot cancels from this stationary equation. If
$q=\mathbb P(x<0)$ is the fraction of blocks that jump in one update,
stationarity gives $\bar v=\overline{\Delta}\,q$, while Eq.~\eqref{eq:xiM} gives
$\overline{x}=f_Y-F-M\bar v$. Stationarity of the second moment gives
\begin{equation}
 \begin{aligned}
 \overline{x^2}={}&\overline{(x-\bar v)^2}\\
 &+2\overline{\Delta}\,\overline{(x-\bar v)\Theta(-x)}
 +\overline{\Delta^2}\,\mathbb P(x<0).
 \end{aligned}
 \label{eq:stationary-second-moment}
\end{equation}
We now specialize to $g(\Delta)=e^{-\Delta}$. For $x>0$, the stationary
equation gives $\rho(x)\propto e^{-x}$, so that
$\overline{x}_{\,x>0}=1$. Since $q=\bar v$, the stable blocks contribute
\begin{equation}
 \overline{x\Theta(x)}=(1-\bar v)\overline{x}_{\,x>0}=1-\bar v
 \label{eq:stable-first-moment}
\end{equation}
to the first moment. With $f_Y=1$, $\overline{\Delta}=1$, and
$\overline{\Delta^2}=2$, Eq.~\eqref{eq:Fc0} gives $F_{c0}=0$, and
Eq.~\eqref{eq:stationary-second-moment} reduces to
\begin{equation}
 (2M-1)\bar v^2+2(F+1-M)\bar v=2F.
 \label{eq:velocity_force_exponential}
\end{equation}

\begin{figure}[ht]
    \centering
    \includegraphics[width=\columnwidth]{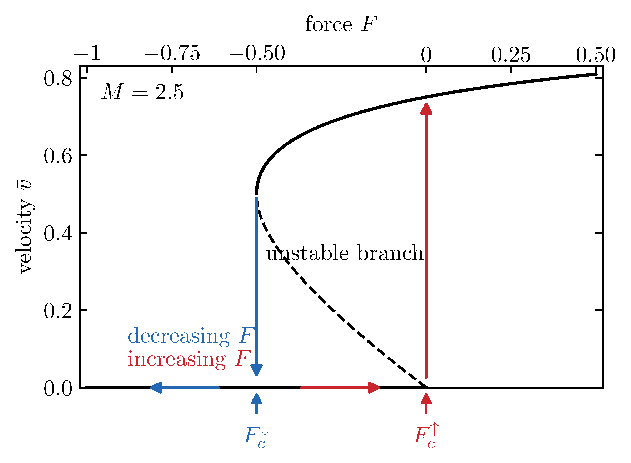}
    \caption{Hysteresis in the force--velocity response for $M=2.5$,
    $f_Y=1$, and $g(\Delta)=e^{-\Delta}$. Upon increasing the force, the
    system jumps from the pinned state to the high-velocity branch at
    $F_c^{\uparrow}=F_{c0}$. Upon decreasing the force, the interface repins at
    $F_c^{\downarrow}$.}
    \label{fig:vf_curve}
\end{figure}

\paragraph{Main results.}

The stationary moving solutions are determined by
Eq.~\eqref{eq:velocity_force_exponential}, whereas the pinned solution has
$\bar v=0$. For $M<1$, depinning is continuous at $F_c=F_{c0}=0$, with
$\bar v\simeq F/(1-M)$; the linear response therefore diverges as
$M\to1^-$. At the marginal value $M=1$, the onset becomes
$\bar v\simeq\sqrt{2F}$, separating continuous depinning from the
discontinuous response found for $M>1$.

For $M>1$, the force--velocity curve becomes reentrant and two distinct
thresholds appear,
\begin{equation}
 F_c^{\downarrow}=-M+\sqrt{2M-1},
 \qquad
 F_c^{\uparrow}=F_{c0}=0.
\end{equation}
For $F<F_c^{\downarrow}$, only the pinned stationary state exists. In the
interval $F_c^{\downarrow}<F<F_c^{\uparrow}$, the pinned state coexists with
a stable high-velocity solution; the two are separated by an unstable
low-velocity branch. When the force is increased quasistatically from a
pinned configuration, motion starts discontinuously at $F_c^{\uparrow}$.
When the force is decreased along the moving branch, the interface repins at
$F_c^{\downarrow}$. The two thresholds therefore delimit the hysteresis loop
shown in Fig.~\ref{fig:vf_curve}.

\section{Elastic driving}
\label{sec:elastic-driving}
\begingroup
We now replace the imposed force by the spring defined in Sec.~I. The common
driving position moves at velocity $V=\dot w$, while $\bar v_t$ denotes the
instantaneous mean velocity of the interface. These velocities coincide only
in a stationary moving state. We focus on $M>1$ and proceed in three steps: the stationary
force--velocity characteristic, the dynamics at finite $V$, and the
quasistatic limit.

\subsection{Stationary force--velocity characteristic}
\label{sec:elastic-stationary}

For $f_Y=1$, the distance to instability under elastic driving is
\begin{equation}
 x_i(t)=1-[h_{\rm CM}(t)-h_i(t)]-k_0[w(t)-h_i(t)]-M\bar v_t.
\end{equation}
At finite driving velocity, its microscopic update is
\begin{equation}
 x_i(t+1)=x_i(t)+(1+k_0)\Delta_i(t)-k_0V
 +M\bar v_t-(1+M)\bar v_{t+1}.
 \label{eq:elastic-update-finite-v}
\end{equation}
In a stationary moving state, $\bar v_t=\bar v_{t+1}=V$, and therefore
\begin{equation}
 x_i(t+1)=x_i(t)+(1+k_0)\bigl[\Delta_i(t)-V\bigr].
\end{equation}
The rescaled variable $y_i=x_i/(1+k_0)$
obeys the same update rule as the distance to instability in the
force-controlled model of Sec.~II at mean velocity $\bar v=V$:
\[
 y_i(t+1)=y_i(t)+\Delta_i(t)-V.
\]
We can therefore use the stationary solution derived there, provided we
identify the force corresponding to the rescaled population. Averaging the
definition of $x_i$, the elastic term $h_{\rm CM}-h_i$ vanishes, while the
mean spring force is $F_{\rm av}$. Hence
$\overline{x}=1-F_{\rm av}-MV$ and
$\overline{y}=\overline{x}/(1+k_0)$. Under force control,
Eq.~\eqref{eq:xiM} instead gives $\overline{x}=1-F-MV$ in a stationary
state with $\bar v=V$. We thus define an auxiliary effective force
$F_{\rm eff}$ so that the mean of the rescaled population takes this same
form:
\begin{equation}
 1-F_{\rm eff}-MV=\overline{y}
 =\frac{1-F_{\rm av}-MV}{1+k_0},
\end{equation}
The physical mean spring force remains $F_{\rm av}$; this correspondence
applies to stationary states. Solving for $F_{\rm eff}$ gives
\begin{equation}
 F_{\rm eff}=\frac{F_{\rm av}+k_0(1-MV)}{1+k_0}.
\end{equation}
Substituting $F_{\rm eff}$ into the constant-force result
Eq.~\eqref{eq:velocity_force_exponential} gives
\begin{equation}
 (2M-1-k_0)V^2
 +2(F_{\rm av}+1-M+2k_0)V
 =2(F_{\rm av}+k_0).
 \label{eq:finite-v-characteristic}
\end{equation}
It reduces to the force-controlled characteristic when $k_0\to0$.
The stationary branch has $0<V<1$, with $V\to1$ only as
$F_{\rm av}\to\infty$.

At $V=0$, Eq.~\eqref{eq:finite-v-characteristic} gives
$F_c^{\uparrow}(k_0)=-k_0$. We define $V_{\rm cr}(k_0)$ as the velocity at
the turning point of the stationary characteristic. Solving
Eq.~\eqref{eq:finite-v-characteristic} for $F_{\rm av}(V)$ and imposing
$dF_{\rm av}/dV=0$ gives
\begin{equation}
 V_{\rm cr}(k_0)=1-\sqrt{\frac{1+k_0}{2M-1-k_0}},
 \label{eq:critical-driving-velocity}
\end{equation}
at the force
\begin{equation}
 F_c^{\downarrow}(k_0)=-M-k_0
 +\sqrt{(1+k_0)(2M-1-k_0)}.
 \label{eq:finite-v-thresholds}
\end{equation}
A positive $V_{\rm cr}$ exists only for
\begin{equation}
 k_0<k_{\rm w},\qquad k_{\rm w}=M-1.
 \label{eq:kw}
\end{equation}
Thus $k_{\rm w}$ marks the onset of velocity weakening. For
$k_0\geq k_{\rm w}$, the stationary characteristic is monotonic; below it,
the characteristic contains a branch with $dF_{\rm av}/dV<0$.

\begin{figure}[t]
 \centering
 \includegraphics[width=0.92\columnwidth]{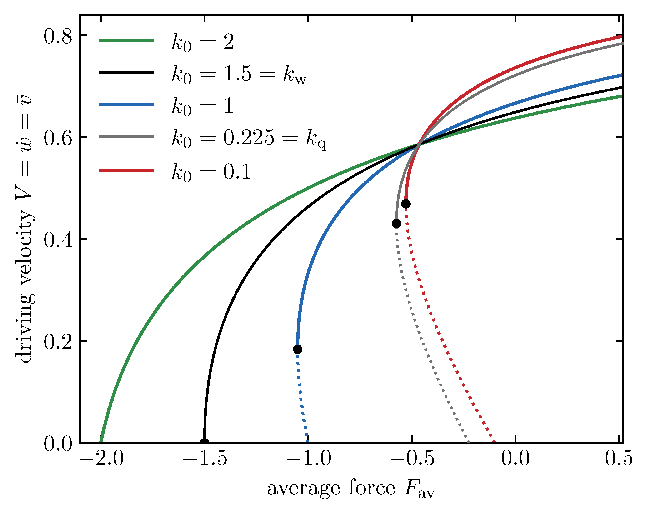}
 \caption{Stationary average-force--velocity characteristics for $M=2.5$.
 Colored curves correspond to representative values of $k_0$; the black and
 gray curves mark $k_0=k_{\rm w}$ and $k_0=k_{\rm q}$, respectively. Dotted
 segments are velocity weakening. Black circles mark $V_{\rm cr}(k_0)$,
 where $dF_{\rm av}/dV=0$; for $k_0=k_{\rm w}$ this point lies at $V=0$.
 In a stationary state, $\bar v=V=\dot w$. Note that all curves pass through
 the common point
 $(F_{\rm av},V)=(1-M(2-\sqrt{2}),\,2-\sqrt{2})$.}
 \label{fig:finite-v-characteristics}
\end{figure}

\subsection{Two dynamical scenarios below
\texorpdfstring{$k_{\rm w}$}{k-w}}
\label{sec:elastic-finite-v}

To determine whether stationary sliding is stable, we study the time
evolution of the density $\rho_t(x)$ of distances to instability in the
thermodynamic limit. From Eq.~\eqref{eq:elastic-update-finite-v},
\begin{align}
 \rho_{t+1}(x)={}&\rho_t(x-A_t)\Theta(x-A_t)\nonumber\\
 &+\int_0^\infty g(\Delta)
 \rho_t\!\left(x-A_t-(1+k_0)\Delta\right)\nonumber\\
 &\qquad\times\Theta\!\left(A_t+(1+k_0)\Delta-x\right)
 \,\mathrm d\Delta,
 \label{eq:finite-v-population-map}
\end{align}
where
\begin{equation}
 A_t=-k_0V+M\bar v_t-(1+M)\bar v_{t+1}.
 \label{eq:finite-v-shift}
\end{equation}
The first term describes blocks that were stable at time $t$ and do not
jump. The integral describes unstable blocks that jump by $\Delta$, drawn
from $g(\Delta)$, increasing their distance to instability by
$(1+k_0)\Delta$. Both populations undergo the common shift $A_t$ due to the
drive, elastic feedback, and stress overshoot. Since the mean jump length
is one, the velocity generated by update $t$ is
\begin{equation}
 \bar v_{t+1}=\int_{-\infty}^{0}\rho_t(x)\,\mathrm dx.
 \label{eq:finite-v-population-velocity}
\end{equation}

We solve these equations numerically at fixed $V$ and compute the average
spring force from
\begin{equation}
 F_{\rm av}(w_t)=1-\int x\rho_t(x)\,\mathrm dx-M\bar v_t.
 \label{eq:finite-v-force}
\end{equation}
The numerical details are given in Appendix~\ref{app:finite-velocity}; here
we discuss the resulting stationary and oscillatory regimes.

For the two cases studied here, population dynamics locates the loss of
stationary sliding at $V_{\rm cr}(k_0)$ within numerical resolution. Above
this velocity the population reaches stationary sliding; below it the force
follows a stick-slip cycle between $F_{\min}(k_0,V)$ and
$F_{\max}(k_0,V)$. Two different limits are found as $V\to0$.

\begin{figure*}[t]
 \centering
 \includegraphics[width=0.94\textwidth]{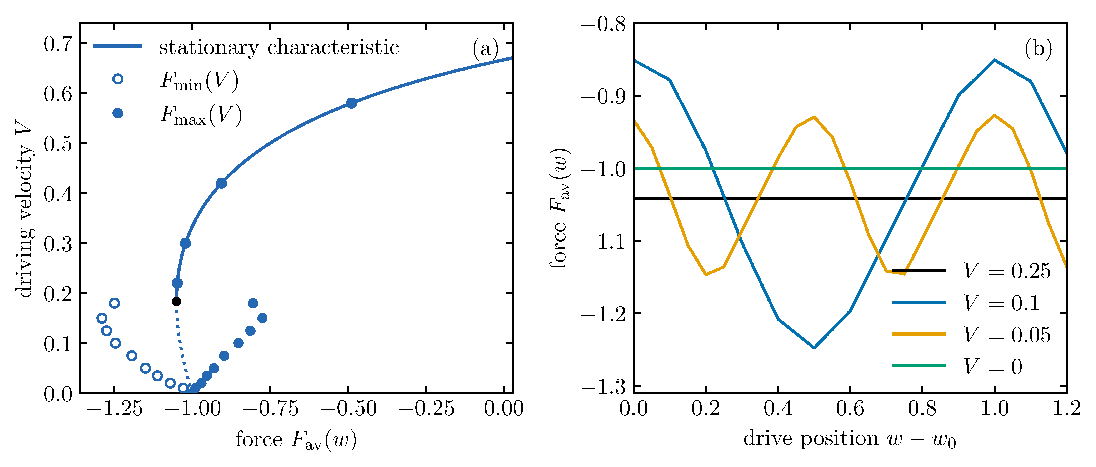}
 \caption{Elastic-driving dynamics for $M=2.5$ and $k_0=1$.
 (a) Stationary characteristic (blue line) and force extrema from population
 dynamics: open circles show $F_{\min}(V)$ and filled circles show
 $F_{\max}(V)$. The dotted segment is velocity weakening and the black circle
 marks $V_{\rm cr}$. (b) Population-averaged force $F_{\rm av}(w)$ versus
 drive position for the velocities indicated in the legend. Horizontal
 trajectories correspond to stationary sliding; the nonhorizontal
 trajectories are stick-slip cycles whose amplitude vanishes as $V\to0$.}
 \label{fig:finite-v-regimes}
\end{figure*}

For $k_0=1$, Fig.~\ref{fig:finite-v-regimes} shows that the oscillation
amplitude decreases as $V$ is reduced towards zero and vanishes in the
quasistatic limit:
\begin{equation}
 \lim_{V\to0}F_{\min}(k_0,V)
 =\lim_{V\to0}F_{\max}(k_0,V)=-k_0.
 \label{eq:intermediate-limit}
\end{equation}

\begin{figure*}[t]
 \centering
 \includegraphics[width=0.94\textwidth]{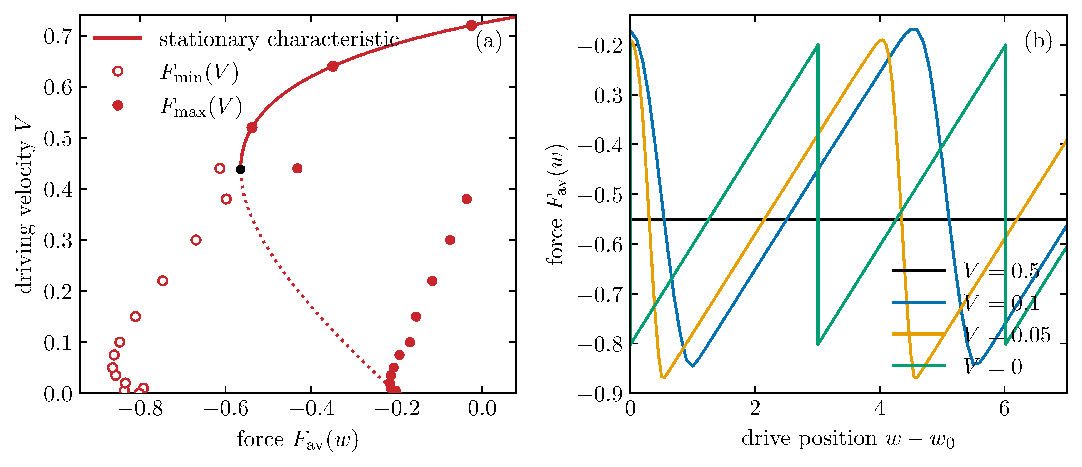}
 \caption{Elastic-driving dynamics for $M=2.5$ and $k_0=0.2$.
 (a) Stationary characteristic (red line) and force extrema from population
 dynamics: open circles show $F_{\min}(V)$ and filled circles show
 $F_{\max}(V)$. The dotted segment is velocity weakening and the black circle
 marks $V_{\rm cr}$. The two extrema remain separated as $V\to0$.
 (b) Population-averaged force $F_{\rm av}(w)$ versus drive position. The
 green $V=0$ curve is the reconstructed quasistatic reference cycle
 (Appendix~\ref{app:finite-velocity}); its force amplitude remains finite.}
 \label{fig:force-versus-w}
\end{figure*}

For $k_0=0.2$, Fig.~\ref{fig:force-versus-w} instead shows a force cycle that
retains a finite amplitude:
\begin{equation}
 \begin{aligned}
  \lim_{V\to0}F_{\max}(k_0,V)&=-k_0,\\
  \lim_{V\to0}F_{\min}(k_0,V)&=F_{\min}(k_0)<-k_0.
 \end{aligned}
 \label{eq:small-stiffness-limit}
\end{equation}
The quasistatic analysis below identifies the stiffness that separates these
two behaviors.

\subsection{Quasistatic limit and avalanche statistics}
\label{sec:quasistatic}

\begin{figure*}[t]
 \centering
 \begin{subfigure}[t]{0.46\textwidth}
  \centering
  \includegraphics[width=\linewidth]{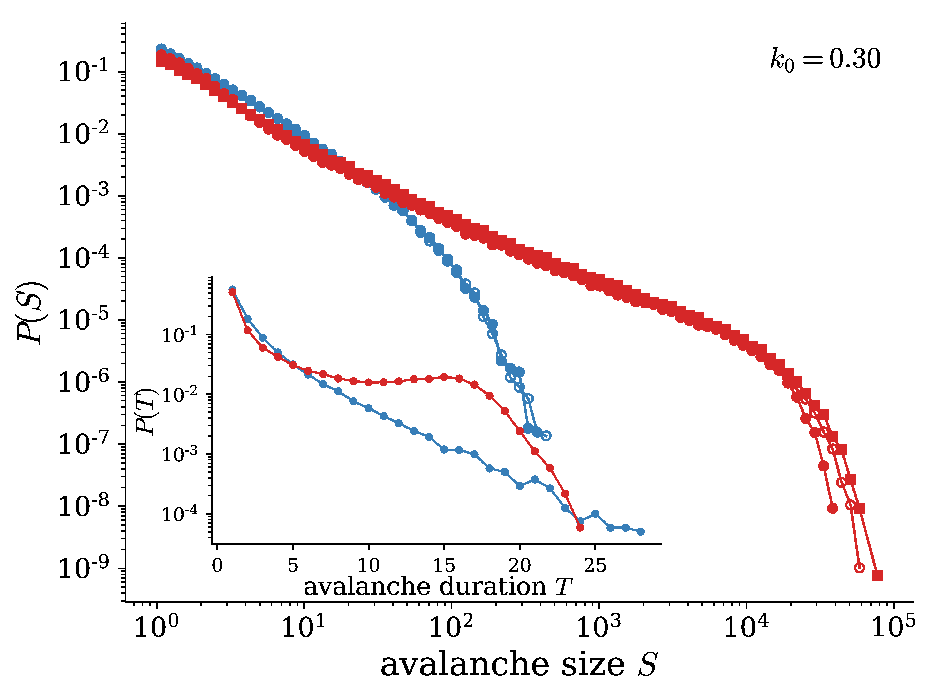}
  \caption*{(a)}
 \end{subfigure}
 \hfill
 \begin{subfigure}[t]{0.46\textwidth}
  \centering
  \includegraphics[width=\linewidth]{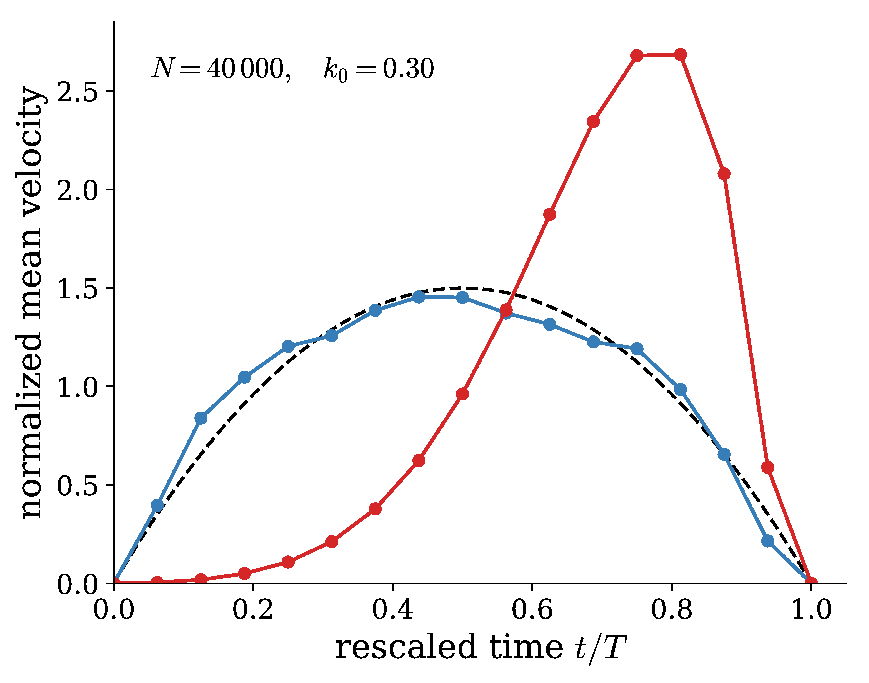}
  \caption*{(b)}
 \end{subfigure}
 \caption{Quasistatic avalanches at $k_0=0.30$.
 (a) Size distribution $P(S)$ for the overdamped reference ($M=0$, blue) and
 the overshoot model ($M=2.5$, red); the inset shows $P(T)$ at
 $N=4\times10^4$. Filled circles, open circles, and filled squares denote
 $N=4\times10^4$, $8\times10^4$, and $1.6\times10^5$, respectively; the
 overdamped reference is shown for the first two sizes.
 (b) Unit-area mean avalanche shapes at fixed duration $T=15$ and
 $N=4\times10^4$. Blue and red circles denote $M=0$ and $M=2.5$; the black
 dashed line is the mean-field critical parabola
 $6(t/T)(1-t/T)$~\cite{PapanikolaouEtAl2011}.}
 \label{fig:quasistatic-avalanches}
\end{figure*}

In the quasistatic protocol, $w$ is increased until the first block becomes
unstable and is then held fixed until all activity has stopped.

We first construct a reference pinned population in which each block
occupies the first sticky patch beyond its stability
threshold~\cite{JaglaLandesRosso2014,OzawaBerthierBiroliRossoTarjus2018}.
Let $r>0$ be the distance from this threshold to the patch. For
exponentially distributed patch spacings of unit mean, the probability of
finding no patch over a distance $r$ is $e^{-r}$. The distance to the first
patch therefore has density $p(r)=e^{-r}$. Since the distance to
instability increases with the block position with slope $1+k_0$, we have
$x=(1+k_0)r$. Changing variables gives
\begin{equation}
 \rho_{\rm st}(x)=\frac{1}{1+k_0}e^{-x/(1+k_0)}\Theta(x).
 \label{eq:stationary-quasistatic}
\end{equation}
This distribution describes the reference pinned state whose response we
study below. Its mean is $\overline{x}=1+k_0$. In a pinned state the
overshoot vanishes, and averaging the definition of $x_i$ gives
$\overline{x}=1-F_{\rm av}$. The corresponding force is therefore
\begin{equation}
 F_{\rm av}^{\rm st}=-k_0.
\end{equation}
This coincides with the quasistatic force in
Eq.~\eqref{eq:intermediate-limit}.

We now trigger an infinitesimal avalanche and follow its activity while
keeping $w$ fixed. We denote by $u_t=h_{\rm CM}(t)-h_{\rm CM}(t-1)$ the
small mean displacement of the interface between times $t-1$ and $t$.
The center of mass advances by $u_t$, increasing the elastic force on each
block that remains immobile by the same amount. The overshoot changes by
$M(u_t-u_{t-1})$. The total force change acting on these blocks is thus
\begin{equation}
 \delta f_t=u_t+M(u_t-u_{t-1})=(1+M)u_t-Mu_{t-1}.
 \label{eq:quasistatic-force-increment}
\end{equation}
If $\delta f_t>0$, the stable blocks initially lying within
$0<x<\delta f_t$ become unstable. To first order, their fraction is
$\rho_{\rm st}(0^+)\delta f_t$. If $\delta f_t\leq0$, no additional stable
blocks are destabilized.

These newly unstable blocks jump during the next update. Since the mean
jump length is one, their fraction equals the mean displacement they
generate. Using $\rho_{\rm st}(0^+)=1/(1+k_0)$, we obtain
\begin{align}
 u_{t+1}
 &=\rho_{\rm st}(0^+)[\delta f_t]_+ +O(u^2)\nonumber\\
 &=\frac{\bigl[(1+M)u_t-Mu_{t-1}\bigr]_+}{1+k_0}+O(u^2),
 \label{eq:one-sided-recurrence}
\end{align}
where $[z]_+=\max(z,0)$. Only an $O(u)$ fraction of blocks jump, and only
an $O(u)$ fraction of these remain unstable. Their contribution is therefore
$O(u^2)$.

As long as $\delta f_t>0$, the activity therefore follows the linear
recurrence
\begin{equation}
 u_{t+1}=\frac{1+M}{1+k_0}u_t
 -\frac{M}{1+k_0}u_{t-1}.
 \label{eq:linear-activity-recurrence}
\end{equation}
The positive-part prescription in Eq.~\eqref{eq:one-sided-recurrence} is
essential when interpreting this recurrence. If its right-hand side becomes
nonpositive, the physical activity vanishes within the linear approximation
and the avalanche stops.

In matrix form, the recurrence reads
\begin{equation}
 \begin{pmatrix}u_{t+1}\\u_t\end{pmatrix}
 =\underbrace{
 \begin{pmatrix}
 \dfrac{1+M}{1+k_0} & -\dfrac{M}{1+k_0}\\[4pt]
 1 & 0
 \end{pmatrix}}_{\mathsf L}
 \begin{pmatrix}u_t\\u_{t-1}\end{pmatrix}.
 \label{eq:activity-evolution-matrix}
\end{equation}
The eigenvalues of $\mathsf L$ determine the growth or decay of the linear
modes. The characteristic equation $\det(\mathsf L-\lambda I)=0$ gives
\begin{equation}
 (1+k_0)\lambda^2-(1+M)\lambda+M=0,
 \label{eq:activity-characteristic-polynomial}
\end{equation}
with roots
\begin{equation}
 \lambda_{\pm}
 =\frac{1+M\pm\sqrt{(M-1)^2-4Mk_0}}
 {2(1+k_0)}.
 \label{eq:activity-eigenvalues}
\end{equation}
The discriminant vanishes at
\begin{equation}
 k_{\rm q}=\frac{(M-1)^2}{4M}.
 \label{eq:kq}
\end{equation}

For $k_0<k_{\rm q}$, the two eigenvalues are real and
$\lambda_+>1$. The activity can grow without changing sign and produce a
macroscopic avalanche. The boundary $k_0=k_{\rm q}$ belongs to the same
side: the repeated eigenvalue $2M/(M+1)$ is still larger than one. For
$k_{\rm q}<k_0<k_{\rm w}$, the eigenvalues are complex, with
\begin{equation}
 |\lambda_{\pm}|=\sqrt{\frac{M}{1+k_0}}>1.
 \label{eq:activity-eigenvalue-modulus}
\end{equation}
Writing $\lambda_\pm=R e^{\pm i\theta}$, the real solution of the signed
recurrence has the form
\[
 u_t=C R^t\cos(t\theta+\phi),
\]
where $C$ and $\phi$ are set by the initial perturbation. The factor
$R^t$ amplifies the envelope, while the oscillating factor eventually
changes sign. At the first step for which
$(1+M)u_t-Mu_{t-1}\leq0$, the positive-part prescription in
Eq.~\eqref{eq:one-sided-recurrence} sets $u_{t+1}=0$. Physically, the loss
of the previous overshoot now outweighs the force increase due to the new
motion, so no additional stable blocks are destabilized. Within the linear
approximation, the next update then gives
\[
 u_{t+2}
 =\left[\frac{(1+M)\,0-Mu_t}{1+k_0}\right]_+=0.
\]
Two consecutive activities are therefore zero, and every subsequent update
also gives zero. Since $w$ is held fixed during the avalanche, there is no
drive increment to restart the motion.

For fixed $k_{\rm q}<k_0<k_{\rm w}$, the first sign change occurs after a
finite number of steps, so the amplification accumulated before arrest is
finite. An infinitesimal seed therefore remains within the linear regime
until the avalanche stops. As $k_0$ approaches $k_{\rm q}$ from above,
$\theta$ tends to zero, delaying the sign change and allowing much larger
transient amplification.

Finally, for $k_0>k_{\rm w}$ the eigenvalue modulus is smaller than one and
perturbations decay. Thus $k_{\rm w}$ marks the
onset of velocity weakening, whereas $k_{\rm q}$ marks the onset of a
quasistatic macroscopic runaway.

For the exponential model, the upper endpoint at the boundary is
$F^*\equiv F_{\max}(k_{\rm q})=-k_{\rm q}$. At and below $k_{\rm q}$,
quasistatic loading reaches $F_{\max}(k_0)=-k_0$ and triggers a genuine king.
The resulting spring unloading ends at a dynamically selected force
$F_{\min}(k_0)<-k_0$, which we do not analyze further here.

We finally examine the finite avalanches in the intermediate interval
$k_{\rm q}<k_0<k_{\rm w}$. We use the extensive activity
$v_{t+1}=\sum_i\Delta_i(t)=Nu_{t+1}$, distinct from the mean interface
velocity; below we relabel the update index as $t$. The duration $T$ is the
number of parallel updates before the activity returns to zero, and the
avalanche size is
\begin{equation}
 S=\sum_{t=1}^{T}v_t.
\end{equation}
Simulation and conditioning details are given in
Appendix~\ref{app:numerics}.

Figure~\ref{fig:quasistatic-avalanches}(a) shows that the overshoot model
develops a broad excess of large events beyond the cutoff of the overdamped
reference, while $P(T)$ develops a shoulder at intermediate durations. The
broad excess is nearly $N$-independent over the common resolved range,
although the far-tail cutoff still shows finite-size dependence. At fixed
duration, panel
(b) shows a second difference: the overdamped mean shape is close to the
symmetric mean-field critical parabola~\cite{PapanikolaouEtAl2011}, whereas
the overshoot model has a strongly
asymmetric profile with a late maximum. These large events are consistent
with the \emph{ghost-king} scenario suggested by the stability analysis. Their
activity is transiently amplified because
$|\lambda_\pm|>1$, but the oscillatory mode reaches the absorbing boundary
before the event becomes system spanning. At and below $k_{\rm q}$, the
nonoscillatory runaway instead produces genuine kings and a macroscopic
quasistatic force drop.

\color{white}
\begin{widetext}
\begin{center}
 \color{black}
 {\small
 \renewcommand{\arraystretch}{1.24}
 \setlength{\tabcolsep}{5pt}
 \begin{tabular}{@{}
  >{\raggedright\arraybackslash}p{0.17\textwidth}
  >{\centering\arraybackslash}p{0.22\textwidth}
  >{\centering\arraybackslash}p{0.26\textwidth}
  >{\centering\arraybackslash}p{0.22\textwidth}@{}}
 \toprule
 \textbf{Property} &
 \textbf{Small stiffness}\newline $k_0\leq k_{\rm q}$ &
 \textbf{Intermediate stiffness}\newline $k_{\rm q}<k_0<k_{\rm w}$ &
 \textbf{Large stiffness}\newline $k_0>k_{\rm w}$ \\
 \midrule
 Stationary characteristic &
 Velocity weakening &
 Velocity weakening &
 Velocity strengthening \\
 Finite-$V$ response &
 Stick-slip for $0<V<V_{\rm cr}$ &
 Stick-slip for $0<V<V_{\rm cr}$ &
 Stable stationary sliding for $0<V<1$ \\
 Quasistatic force response &
 Finite force cycle &
 Cycle amplitude vanishes; $F_{\rm av}\to-k_0$ &
 $F_{\rm av}\to-k_0$ \\
 Pinned-state perturbation &
 Nonoscillatory runaway &
 Oscillatory amplification followed by arrest &
 Decay \\
 Quasistatic avalanches &
 Genuine kings (system-spanning events) &
 Ghost kings (large but finite events) &
 Finite avalanches \\
 \bottomrule
 \end{tabular}}
 \captionof{figure}{Summary of the three elastic-driving regimes for $M>1$,
 with $k_{\rm q}=(M-1)^2/(4M)$ and $k_{\rm w}=M-1$. The boundary
 $k_0=k_{\rm q}$ belongs to the genuine-king regime, whereas
 $k_0=k_{\rm w}$ is marginal.}
 \label{fig:regime-map}
\end{center}
\end{widetext}
\endgroup
\section{Discussion}
\begingroup

Figure~\ref{fig:regime-map} collects the three regimes established above.
Under force control,
the reentrant force--velocity characteristic and runaway motion are two
manifestations of the same discontinuous depinning transition. Elastic
feedback separates their onset into two distinct stiffness thresholds.
The value $k_{\rm w}=M-1$ is read from the stationary force--velocity
characteristic: below it, a velocity-weakening branch exists.  By contrast,
$k_{\rm q}$ is determined by the response of a pinned state to a small
activity perturbation. It determines whether the amplified perturbation is
arrested when the activity first returns to zero or develops into a
macroscopic avalanche. This distinction is the origin of the intermediate
regime and is the main conceptual result of this work.

The intermediate regime displays stick-slip at small finite $V$, but the
force cycle vanishes as $V\to0$ and the interface-averaged spring force
$F_{\rm av}$ converges to its pinned-state value $-k_0$, even though individual
avalanches can undergo strong transient amplification. The measured excess
of large avalanches, shoulder in the duration distribution, and asymmetric
mean shape are consistent with this transient amplification. We call these
large but finite events \emph{ghost kings}, reserving \emph{genuine kings} for the macroscopic
events that unload the system at and below $k_{\rm q}$.

Large but non-system-spanning avalanches also occur in related models with
transient weakening~\cite{JaglaLandesRosso2014}. In the finite-dimensional
short-range model of Ref.~\cite{PetrilloJaglaLippielloRosso2026}, where these
events were termed \emph{ghost kings}, initially supercritical avalanches
propagate ballistically but are arrested at a characteristic size that does
not grow with the system. Here the fully connected geometry eliminates spatial
propagation: the transient instead follows an amplified oscillatory mean-field
mode before returning to the absorbing state. In both cases, an avalanche
starts with supercritical growth but fails to become system spanning.

The force-controlled analysis of stress overshoots
\cite{SchwarzFisher2001,SchwarzFisher2003} identifies coexistence between
stationary moving and pinned states for
$F_c^{\downarrow}<F<F_c^{\uparrow}$. Under elastic driving, the spring force
decreases during a macroscopic avalanche. The stopping force $F_{\min}(k_0)$
is selected by the nonstationary avalanche dynamics and should not be
identified with the turning force $F_c^{\downarrow}(k_0)$ of the stationary
moving branch.

The aging model of Ref.~\cite{PereyraAponteJagla2026} also displays velocity
weakening and stick-slip, but with a different phase diagram. In mean field,
the cycles are mediated by king avalanches, whereas in two dimensions they
consist of alternating periods of high and low activity without system-spanning
events. Thus, macroscopic stress cycles do not necessarily imply king
avalanches.

The relation between instability and macroscopic failure is not one-to-one.
Related questions arise in frictional rupture and collective damage
localization \cite{deGeusPopovicJiRossoWyart2019,deGeusWyart2022,
DansereauEtAl2019}. In the nucleation scenario for brittle yielding, a
perturbation must exceed a finite critical size before it runs away
\cite{OzawaBerthierBiroliRossoTarjus2018,PopovicDeGeusWyart2018}; in an
alternative scenario, the homogeneous state is linearly unstable and an
arbitrarily small perturbation ultimately produces macroscopic localization
\cite{BarlowCochranFielding2020}. For ghost kings, the eigenvalues describe
the linear evolution only while the activity remains positive. In the intermediate
regime, the avalanche stops when the loss of the previous overshoot outweighs
the force increase due to new motion. Within the linear approximation, once
the activity reaches zero, it remains zero because the drive is held fixed.
Thus, eigenvalues with modulus greater than one can amplify a finite avalanche
without producing a macroscopic runaway.

Our results rely on fully connected interactions, which suppress spatial
propagation, and exponential sticky-patch spacings, which make the calculations
tractable. The thresholds $k_{\rm w}$ and $k_{\rm q}$ are therefore model
dependent. It remains to establish whether finite-dimensional ghost kings
share this amplification mechanism and whether short-range systems exhibit
analogous force oscillations at finite velocity.
\endgroup

\begin{acknowledgments}
We thank E. Jagla, E. Lippiello, and G. Petrillo for helpful discussions.
This work was supported by the French National Research Agency (ANR) through
the DISCREEP project (ANR-23-CE30-0031-04) and by the CNRS International
Research Project ``Statistical Physics of Materials''.
\end{acknowledgments}

\appendix
\section{Numerical solution of the population dynamics}
\label{app:finite-velocity}
The finite-velocity results in Figs.~\ref{fig:finite-v-regimes}
and~\ref{fig:force-versus-w} are obtained by iterating
Eqs.~\eqref{eq:finite-v-population-map}--\eqref{eq:finite-v-population-velocity}
on a uniform grid of cell centers covering $x\in[-30,40]$. We evolve the
probability density directly, without sampling individual blocks.

At each update, we first compute $\bar v_{t+1}$ and $A_t$ from the current
density. To evaluate the jump contribution, we use the probability density
of the displacement $z=(1+k_0)\Delta$,
\begin{equation}
 K_{k_0}(z)=\frac{e^{-z/(1+k_0)}}{1+k_0}\Theta(z).
 \label{eq:population-kernel}
\end{equation}
The unstable part of the density is convolved with this kernel, using
weights obtained by integrating $K_{k_0}$ over each displacement bin.
We then add the stable part and apply the common shift $A_t$ by linear
interpolation. The density is taken to vanish outside the numerical domain
and is normalized after each update.

Runs start from the density in Eq.~\eqref{eq:stationary-quasistatic}, with
the velocity history initialized by $\bar v_0=V$. The force extrema in
panels (a) use $\delta x=0.02$ and $8000$ updates, with the first $4000$
discarded. The trajectories in panels (b) use $\delta x=0.01$ and $12000$
updates, with the first $5000$ discarded. The force is evaluated using
Eq.~\eqref{eq:finite-v-force}; its minimum and maximum over the retained
updates give $F_{\min}(k_0,V)$ and $F_{\max}(k_0,V)$. Consecutive updates
are separated by $V$ along the horizontal axis, with the origin $w_0$
chosen separately for each displayed trajectory.

The curves labelled $V=0$ represent the quasistatic limit and are
constructed separately. For Fig.~\ref{fig:force-versus-w}, we first compute
a single avalanche at fixed $w$. We start from the stable density in
Eq.~\eqref{eq:stationary-quasistatic} and multiply all its grid values by
$1-\epsilon$, removing a total probability mass $\epsilon$. We place this
mass in the first cell below $x=0$ by adding $\epsilon/\delta x$ to its
density. The total probability remains one, but a fraction $\epsilon$ of
the population is now unstable.

With $\bar v_0=0$, the unstable fraction jumps at the first update and
generates $\bar v_1=\epsilon$, since the mean jump length is one. We then
iterate the population map until the activity falls below $10^{-12}$.
The spring force decreases by $k_0\bar v_{t+1}$ at each
update, and its final value defines $F_{\min}$. For the reference in
Fig.~\ref{fig:force-versus-w}, we use $\epsilon=10^{-8}$,
$x\in[-100,400]$, and $\delta x=0.005$, so that the seed occupies the cell
$[-0.005,0]$. This gives $F_{\min}\simeq-0.80177$ at $k_0=0.2$.

We plot the calculated avalanche as a vertical force drop from $-k_0$ to
$F_{\min}$. We then add a straight elastic reloading branch of slope $k_0$
up to $-k_0$ and repeat these two segments to draw the reference cycle.
In Fig.~\ref{fig:finite-v-regimes}, the quasistatic
reference is simply the horizontal line $F_{\rm av}=-k_0$, since the
macroscopic force drop vanishes in this regime.

\section{Quasistatic avalanche simulations}
\label{app:numerics}
We simulate the quasistatic dynamics directly for $N$ blocks, with $f_Y=1$
and independent exponential sticky-patch spacings of unit mean.

From a stable configuration, the drive is advanced to the next instability by
\begin{equation}
 x_i\leftarrow x_i-\min_jx_j.
\end{equation}
It is then held fixed. At parallel update $t$, every block in
$\mathcal U_t=\{i:x_i(t)\leq0\}$ draws an independent jump $\Delta_i(t)$
from $g(\Delta)=e^{-\Delta}$; for all other blocks $\Delta_i(t)=0$. With
\begin{equation}
 v_{t+1}=\sum_i\Delta_i(t),
 \qquad
 \bar v_{t+1}=\frac{v_{t+1}}{N},
\end{equation}
the distances to instability evolve as
\begin{equation}
 x_i(t+1)=x_i(t)+(1+k_0)\Delta_i(t)
 +M\bar v_t-(1+M)\bar v_{t+1}.
 \label{eq:direct-update}
\end{equation}
Here $\bar v_0=0$ at the start of each avalanche. Updates continue until
$\mathcal U_t$ is empty; $T$ and $S$ are then measured as defined in the main
text.

The data in Fig.~\ref{fig:quasistatic-avalanches} are measured at $k_0=0.30$
for $M=0$ and $M=2.5$. Before recording events, we discard complete
avalanches until every block has toppled at least once. For exponential
spacings, this full-renewal criterion removes the dependence on the initial
configuration. We then record $1.2\times10^5$ avalanches for each displayed
pair $(M,N)$. The $M=0$ data use $N=4\times10^4$ and $8\times10^4$, while
the $M=2.5$ data also include $N=1.6\times10^5$. Size distributions are
evaluated in common logarithmic bins and duration probabilities at integer
$T$; points with fewer than five events are omitted. Mean shapes are
conditioned on $T=15$ at $N=4\times10^4$, padded by one zero at each end,
averaged as functions of $t/T$, and normalized to unit area. The reference
curve is the mean-field critical parabola of
Ref.~\cite{PapanikolaouEtAl2011}. With $s=t/T$, their unit-height form
$4s(1-s)$ becomes $6s(1-s)$ under our unit-area normalization,
$\int_0^1 6s(1-s)\,ds=1$.
\clearpage
\bibliographystyle{journal-full}
\bibliography{biblio}
\end{document}